\documentclass[12pt,a4paper,final]{iopart}
\usepackage{iopams}
\usepackage{graphicx}
\usepackage[breaklinks=true,colorlinks=true,linkcolor=blue,urlcolor=blue,citecolor=blue]{hyperref}

\begin{document}

\title{Composite localized modes in discretized spin-orbit-coupled
Bose-Einstein condensates}
\author{Petra P. Beli\v cev$^{1}$,Goran Gligori\'{c}$^{1}$, Jovana Petrovic$%
^{1}$, Aleksandra Maluckov$^{1}$, Ljup\v{c}o Had\v{z}ievski$^{1}$ and Boris
A. Malomed$^{2}$}

\begin{abstract}
We introduce a discrete model for binary spin-orbit-coupled (SOC)
Bose-Einstein condensates (BEC) trapped in a deep one-dimensional optical
lattice. Two different types of the couplings are considered, with spatial
derivatives acting inside each species, or between the species. The discrete
system with inter-site couplings dominated by the SOC, while the usual
hopping is negligible, \textit{emulates} condensates composed of extremely
heavy atoms, as well as those with opposite signs of the effective atomic
masses in the two components.\ Stable localized composite states of miscible
and immiscible types are constructed. The effect of the SOC on the
immiscibility-miscibility transition in the localized complexes, which
emulates the phase transition between insulating and conducting states in
semiconductors, is studied.
\end{abstract}


\address{$^1$Vin\v{c}a Institute of Nuclear Sciences, University of Belgrade, P. O. B.
522,11001 Belgrade, Serbia}
\address{$^2$Department of Physical Electronics, School of Electrical Engineering,
Faculty of Engineering, Tel Aviv University, Tel Aviv 69978,
Israel} \ead{petrab@vin.bg.ac.rs}


\vspace{2pc} \noindent\textit{Keywords}: spin-orbit coupling, two component
BEC, miscibility-immiscibility transition, discrete soliton complexes
%\submitto{\JPB}
%\maketitle

\section{Introduction}

Recently, the use of ultracold quantum gases -- atomic Bose-Einstein
condensates (BEC) and fermion gases alike -- for simulating fundamental
effects originating in condensed-matter physics has drawn much interest \cite%
{Lewenstein}. One of these effects is the spin-orbit coupling (SOC) which
links the electron spin to its motion in semiconductors. The SOC of the
Dresselhaus \cite{Dresselhaus} and Rashba \cite{Rashba} types plays a major
role in many phenomena and applications, including spin and anomalous Hall
effects \cite{1}, topological insulators \cite{2}, spintronics \cite{3,8},
spin-based quantum computations \cite{sup}, etc. In contrast to the complex
situations found in solids, the ``synthetic" SOC, induced by appropriate
laser illumination of atomic gases in the combination with a magnetic field,
can be precisely controlled in the experiment \cite%
{Spielman-2011,socbec,nature49,China}. Furthermore, unlike the electron
spin, the pseudo-spin of laser-dressed atoms is not constrained by
fundamental symmetries, which gives rise to a variety of settings
unavailable in solids, such as new exotic superfluids \cite{fluid}. The
theoretical description of SOC effects in BEC is also much simpler than in
solids \cite{Spielman-2011,theory-SOC,Basque}. Therefore, studies of the SOC
in BEC, as well as in fermionic gases \cite{7,socfermion}, have become a
vast research area, see Ref. \cite{review-SOC} for a brief review.

SOC in solid state is manifested by electrons moving in static electric
fields \cite{9,10,11}. The coupling results from the Zeeman interaction
between the magnetic moment of the electron, aligned with its spin, and the
magnetic field appearing in the reference frame moving along with the
electron. To link this setting to the atomic gas, the electron's spinor wave
function is mapped into a pseudo-spinor mean-field wave function of the
binary BEC in $^{87}$Rb, which contains atoms in two different states
``dressed" by appropriate laser fields. In particular, making use of the
electronic ground-state $5S_{1/2},F=1$ and $F=2$ hyperfine manifolds, one
can start with the set of four states, $\left\vert F,m_{F}\right\rangle
=\left\vert 2,0\right\rangle ,\left\vert 1,-1\right\rangle ,\left\vert
1,0\right\rangle ,\left\vert 2,1\right\rangle $, which are cyclically
coupled by four strong fields directed along diagonals in the $\left(
x,y\right) $ plane, with the magnetic field applied along $z$. As shown in
\cite{Spielman-2011}, by means of an appropriate unitary transformation,
which dresses the original atomic states, this setting may be described by a
combination of the Rashba and Dresselhaus Hamiltonians, with independently
adjustable coefficients, $\alpha $ and $\beta $, acting on a set of two
dressed states, that emulate the spin-up and spin-down polarizations of
electrons in the semiconductor:%
\begin{equation}
\hat{H}_{1}=\alpha \left( \sigma _{x}\hat{p}_{y}-\sigma _{y}\hat{p}%
_{x}\right) +\beta \left( \sigma _{x}\hat{p}_{y}+\sigma _{y}\hat{p}%
_{x}\right) ,  \label{H1}
\end{equation}%
where $\sigma _{x,y}$ are the Pauli matrices, and $\hat{p}_{x,y}$ are
operators of the respective momentum components.

In the seminal experimental work \cite{socbec}, a different setting was
implemented in the condensate of $^{87}$Rb: a pair of Raman laser beams with
wavelength $\lambda $, shone along the same diagonal directions in the $%
\left( x,y\right) $ plane and with the same vertical orientation of the
magnetic field as mentioned above, were used to couple two states, $%
\left\vert \psi ^{+}\right\rangle =\left\vert F=1,m_{F}=0\right\rangle $ and
$\left\vert \psi ^{-}\right\rangle =\left\vert F=1,m_{F}=-1\right\rangle $
(the third one, $\left\vert F=1,m_{F}=+1\right\rangle $, which belongs to
the same manifold, is far detuned from this pseudo-spinor set). The original
Hamiltonian of the system includes the kinetic energy of atoms, Zeeman
terms, and inter-component coupling terms modulated by factors $\exp \left(
\pm 4\pi ix/\lambda \right) $. The transformation of this Hamiltonian by
unitary matrix $U=\exp \left( 2\pi i\sigma _{z}x/\lambda \right) $ makes it
possible to eliminate the explicit $x$-dependence, simultaneously generating
the effective SOC term,%
\begin{equation}
\hat{H}_{2}=\gamma \sigma _{z}\hat{p}_{x},  \label{H2}
\end{equation}%
due to the noncommutivity of $U$ and term $\hat{p}_{x}^{2}/\left( 2m\right) $
in the kinetic-energy operator. Further, the pseudo-spin rotation can make $%
H_{2}$ equivalent to the combination of the Rashba and Dresselhaus couplings
with equal strength, i.e., a particular case of the generic Hamiltonian (\ref%
{H1}).

The interplay of the SOC with intrinsic collisional nonlinearity of the
pseudo-spinor BEC was recently considered too, in many theoretical works. In
particular, it was demonstrated the new ground-state phases (stripes, phase
separation, etc.) can be created in a such nonlinear two-component systems
\cite{pseudomag}, as well as tricritical points \cite{Tricrit}, solitons of
different types \cite{PRL110,PRL111,brsol,Konotop}, including
two-dimensional solitons with embedded vorticity \cite%
{Fukuoka,Cardoso,Fukuoka2}, and vortex lattices \cite{Saka}. New topological
excitations emerge in spin-orbit-coupled fermionic gases too \cite%
{socfermion}. SOC gives rise to other noteworthy phenomena when combined
with an optical-lattice (OL) potential: flattening of the Bloch potential
\cite{ZhangPRA13}, atomic \textit{Zitterbewegung} \cite{LarsonPRA10}, and
new topological phases \cite{StanescuPRA09}.

Our objective here is to derive a discrete form of the SOC-BEC model for the
pseudo-spinor condensate trapped in a deep OL potential, and then construct
discrete solitons specific to this model. Especially interesting is the
regime in which the inter-site coupling is dominated by SOC, while the usual
hopping may be neglected. It gives rise to a previously unknown discrete
model that makes it possible to emulate the SOC in BEC composed of
infinitely heavy atoms, as well as emulate binary condensates with opposite
signs of effective atomic masses (dynamics of trapped SOC BEC featuring an
infinite effective mass in some special cases was recently addressed in
non-discrete settings both experimentally \cite{China} and theoretically
\cite{Basque}).

We also investigate the immiscibility-miscibility (IM) transition induced by
the SOC in localized pseudo-spin complexes. In this connection, it is
relevant to mention that, in the absence of linear couplings between the
components, the hyperfine states in $^{87}$Rb are immiscible, although being
close to the miscibility threshold \cite{22,22a,22b}. Previously elaborated
proposals to achieve a tunable immiscibility-miscibility transition relied
on the adjustment of the scattering length via the Feshbach resonance,
controlled by magnetic \cite{23,23a} or optical \cite{24,24a} fields, or on
linear coupling between the different atomic species imposed by a
radio-frequency field \cite{transmalomed,miIMT}. The SOC imposes its
specific linear mixing via spatial gradients of the wave functions, thus
offering a new way to induce the immiscibility-miscibility transition.

The SOC model, written in terms of coupled discrete Gross-Pitaevskii
equations (GPEs), is introduced in Section II. Results for discrete
bright-soliton complexes in this system and their stability, as well as for
the effect of the SOC on the immiscibility-miscibility transition in the
binary condensate, are presented in Section III. The paper is concluded by
Section IV.

\section{The model}

\subsection{Basic equations}

We consider the BEC composed of atoms in different hyperfine ground states
corresponding, as said above, to $F=1$ \cite{socbec} or $F=1,2$ \cite%
{Spielman-2011}. The states are coupled by two \cite{socbec} or four \cite%
{Spielman-2011} Raman laser beams shone at angle $\pi /4$ in he $x-y$ plane
with respect to the system's axis $x$, with the dc magnetic field applied
along $z$. These settings induce the SOC in the basis of two dressed states,
which emulate the spin up and down components. In the presence of the OL
potential, $V_{\mathrm{OL}}$, the mean-field dynamics of the BEC is governed
by the GPE written in the spinor form:
\begin{equation}
i\hbar \frac{\partial \Psi }{\partial t}=\left[ \frac{\hat{p}^{2}}{2m}+V_{%
\mathrm{OL}}+\hat{H}_{\mathrm{SOC}}+\hat{H}_{\mathrm{mix/spl}}+H_{\mathrm{int%
}}\right] \Psi ,  \label{eq00}
\end{equation}%
where $\Psi =(\psi ^{+},\psi ^{-})^{T}$ is the normalized spinor wave
function in the dressed-state representation, and $\hat{p}^{2}/\left(
2m\right) $ represents the kinetic energy of the quasi-1D condensate.

According to the above discussion, the Hamiltonian term (\ref{H2}), which
accounts for the SOC in the setting similar to that which was experimentally
realized in \cite{socbec}, amounts to
\begin{equation}
\hat{H}_{\mathrm{SOC}}=-i\frac{\hbar ^{2}}{m}\kappa \sigma _{z}\frac{%
\partial }{\partial x},  \label{HSrd}
\end{equation}%
where $\kappa $ is the SOC strength. In this case, we assume that the
condensate is subject to tight confinement in the $y$ and $z$ directions,
hence $\hat{p}^{2}$ in Eq. (\ref{eq00}) may be replaced by $\hat{p}_{x}^{2}$%
. On the other hand, starting from the more general SOC Hamiltonian (\ref{H1}%
), and assuming the tight confinement in the $\left( z,x\right) $ plane,
i.e., keeping only $\hat{p}_{y}$ in $\hat{p}^{2}$ in Eq. (\ref{eq00}) and in
(\ref{H1}), the SOC Hamiltonian amounts to
\begin{equation}
\hat{H}_{\mathrm{SOC}}=-i\frac{\hbar ^{2}}{m}\kappa \sigma _{x}\frac{%
\partial }{\partial y},  \label{HSro}
\end{equation}%
with respective strength $\kappa $.

The Hamiltonian term (\ref{HSrd}) implies that the spatial derivatives
accounting for the SOC act inside of each dressed state (we call this system
an intra-SOC one), while the term (\ref{HSro}) couples different dressed
states by the derivatives (to be called the inter-SOC system). In the former
and latter cases, the unconfined coordinate is denoted as $x$ and $y$,
respectively. The commutation of the SOC terms in the Hamiltonian with the
respective coordinate gives rise to what may be considered as anomalous
velocity, and effects produced by the SOC terms, such as the
immiscibility-miscibility transition (see below) may be accordingly
understood as mixing caused by this velocity.

Further, the linear-mixing or splitting terms of the Hamiltonian, which are
induced by the magnetic field, are represented, severally, by $\hat{H}_{%
\mathrm{mix}}=\hbar \Omega \sigma _{x}$ and $\hat{H}_{\mathrm{spl}}=\hbar
\Omega \sigma _{z}$ in the intra- and inter-SOC systems, where $\Omega $ is,
respectively, the Rabi coupling or Zeeman splitting between the two
wave-function components. Lastly, $H_{\mathrm{int}}$ accounts for the
collisional intra- and inter-component interactions in the BEC.

The fragmentation of the condensate in the deep OL potential into droplets
coupled by tunnelling across barriers separating local potential wells leads
to the replacement of the continuous GPEs by their discrete counterparts,
which can be performed in essentially the same way (using the tight-binding
approximation) as it was elaborated for the single-component BEC \cite%
{BECinOL,PR463}. Thus, adopting the units where $\hbar =1,\,m=1$, we arrive
at a system of two discrete GPEs for the (pseudo-) spinor wave function, $%
\Psi _{n}=(\psi _{n}^{+},\psi _{n}^{-})^{T}$, where $n$ is the discrete
coordinate, which replaces $x$ in Eq. (\ref{HSrd}), or $y$ in Eq. (\ref{HSro}%
). The first system corresponds to the intra-SOC, cf. Eq. (\ref{HSrd}), with
the first-order finite-difference derivatives acting separately on each
component:
\begin{eqnarray}
i\frac{\partial \psi _{n}^{+}}{\partial t} &=&-C(\psi _{n+1}^{+}+\psi
_{n-1}^{+})-i\kappa (\psi _{n+1}^{+}-\psi _{n-1}^{+})+(\gamma _{1}|\psi
_{n}^{+}|^{2}+\beta |\psi _{n}^{-}|^{2})\psi _{n}^{+}-\Omega \psi _{n}^{-}~,
\nonumber \\
i\frac{\partial \psi _{n}^{-}}{\partial t} &=&-C(\psi _{n+1}^{-}+\psi
_{n-1}^{-})+i\kappa (\psi _{n+1}^{-}-\psi _{n-1}^{-})+(\beta |\psi
_{n}^{+}|^{2}+\gamma |\psi _{n}^{-}|^{2})\psi _{n}^{-}-\Omega \psi _{n}^{+}~.
\label{eq2a}
\end{eqnarray}%
The other model corresponds to the inter-SOC, in which the finite-difference
derivatives mix the components, cf. Eq. (\ref{HSro}):
\begin{eqnarray}
i\frac{\partial \psi _{n}^{+}}{\partial t} &=&-C(\psi _{n+1}^{+}+\psi
_{n-1}^{+})+i\kappa (\psi _{n+1}^{-}-\psi _{n-1}^{-})+(\gamma _{1}|\psi
_{n}^{+}|^{2}+\beta |\psi _{n}^{-}|^{2})\psi _{n}^{+}-\Omega \psi _{n}^{+},
\nonumber \\
i\frac{\partial \psi _{n}^{-}}{\partial t} &=&-C(\psi _{n+1}^{-}+\psi
_{n-1}^{-})+i\kappa (\psi _{n+1}^{+}-\psi _{n-1}^{+})+(\beta |\psi
_{n}^{+}|^{2}+\gamma |\psi _{n}^{-}|^{2})\psi _{n}^{-}+\Omega \psi _{n}^{-}~.
\label{eq2b}
\end{eqnarray}%
In either system $\kappa $ is the rescaled SOC strength, $\gamma ,\gamma
_{1}<0$ and $\beta <0$ are, respectively, coefficients of the intra- and
inter-component interactions, which we assume here to be attractive, and $%
\Omega $ is the Rabi-coupling or Zeeman-splitting frequency in Eqs. (\ref%
{eq2a}) and (\ref{eq2b}), respectively. The nonlinearity coefficients are
normalized by setting $\gamma _{1}\equiv -1$. The remaining free parameters
are $\kappa ,\beta ,\gamma $, and $\Omega $, along with the inter-site
hopping coefficient, $C>0$.

It is well known that the discrete equation with repulsive on-site
nonlinearity can be transformed into its counterpart with the attractive
interaction by means of the staggering transformation followed by the
complex conjugation:%
\begin{equation}
\psi _{n}\equiv (-1)^{n}\tilde{\psi}_{n}^{\ast },  \label{stag}
\end{equation}%
where the asterisk stands for the complex conjugate. If the underlying
nonlinearity is repulsive, the transformation will invert the original signs
of $\gamma _{1}$, $\gamma $ and $\beta $, while the sign of the SOC
coefficient, $\kappa $, will not change.

Each system conserves two quantities, that are written here for infinite
lattices: the norm,
\begin{equation}
P=\sum_{n}(|\psi _{n}^{+}|^{2}+|\psi _{n}^{-}|^{2}),  \label{P}
\end{equation}%
and Hamiltonian, which is%
\begin{eqnarray}
H &=&\sum_{n}\left\{ -C\left( \psi _{n}^{+\ast }\psi _{n+1}^{+}+\psi
_{n}^{-\ast }\psi _{n+1}^{-}\right) -i\kappa \left( \psi _{n}^{+\ast }\psi
_{n+1}^{+}-\psi _{n}^{-\ast }\psi _{n+1}^{-}\right) \right.  \nonumber \\
&&\left. +\frac{1}{4}\left( \gamma _{1}\left\vert \psi _{n}^{+}\right\vert
^{4}+\gamma \left\vert \psi _{n}^{-}\right\vert ^{4}\right) +\frac{\beta }{2}%
\left\vert \psi _{n}^{+}\right\vert ^{2}\left\vert \psi _{n}^{-}\right\vert
^{2}-\Omega \psi _{n}^{+\ast }\psi _{n}^{-}\right\} +\mathrm{c.c.}
\end{eqnarray}%
for the intra-SOC system (\ref{eq2a}), and
\begin{eqnarray}
H &=&\sum_{n}\left\{ -C\left( \psi _{n}^{+\ast }\psi _{n+1}^{+}+\psi
_{n}^{-\ast }\psi _{n+1}^{-}\right) +i\kappa \psi _{n}^{+\ast }\left( \psi
_{n+1}^{-}-\psi _{n-1}^{-}\right) \right.  \nonumber \\
&&\left. +\frac{1}{4}\left( \gamma _{1}\left\vert \psi _{n}^{+}\right\vert
^{4}+\gamma \left\vert \psi _{n}^{-}\right\vert ^{4}\right) +\frac{\beta }{2}%
\left\vert \psi _{n}^{+}\right\vert ^{2}\left\vert \psi _{n}^{-}\right\vert
^{2}-\frac{\Omega }{2}\left( \left\vert \psi _{n}^{+}\right\vert
^{2}-\left\vert \psi _{n}^{-}\right\vert ^{2}\right) \right\} +\mathrm{c.c.}
\end{eqnarray}%
for the inter-SOC system (\ref{eq2b}), where $\mathrm{c.c.}$\ stands for the
complex-conjugate expression.

\subsection{Stationary solutions}

Stationary solutions for the above systems are sought for as $\psi
_{n}^{+}=A_{n}\exp {(-i\mu t)}$ and $\psi _{n}^{-}=B_{n}\exp {(-i\mu t)}$,
where $\mu $ is the chemical potential. The set of stationary equations
following from Eqs. (\ref{eq2a}) is
\begin{eqnarray}
\mu A_{n} &=&-C(A_{n+1}+A_{n-1})-i\kappa (A_{n+1}-A_{n-1})+(\gamma
_{1}|A_{n}|^{2}+\beta |B_{n}|^{2})A_{n}-\Omega B_{n},  \nonumber \\
\mu B_{n} &=&-C(B_{n+1}+B_{n-1})+i\kappa (B_{n+1}-B_{n-1})+(\beta
|A_{n}|^{2}+\gamma |B_{n}|^{2})B_{n}-\Omega A_{n},  \label{eq2as}
\end{eqnarray}%
and its counterpart following from Eq. (\ref{eq2b}) is%
\begin{eqnarray}
\mu A_{n} &=&-C(A_{n+1}+A_{n-1})+i\kappa (B_{n+1}-B_{n-1})+(\gamma
_{1}|A_{n}|^{2}+\beta |B_{n}|^{2})A_{n}-\Omega A_{n},  \nonumber \\
\mu B_{n} &=&-C(B_{n+1}+B_{n-1})+i\kappa (A_{n+1}-A_{n-1})+(\beta
|A_{n}|^{2}+\gamma |B_{n}|^{2})B_{n}+\Omega B_{n},  \label{eq2bs}
\end{eqnarray}%
Our aim is to find stable discrete solitons for both systems, with the
intra- and inter-SOC. To this end, stationary equations (\ref{eq2as}) and (%
\ref{eq2bs}) were solved by means of a numerical algorithm based on the
modified Powell minimization method \cite{mi1D}. Stability of so found
discrete solitons was checked, in the framework of the linear stability
analysis, by numerically solving the corresponding eigenvalue problem for
modes of small perturbations. Finally, Eqs. (\ref{eq2a}) and (\ref{eq2b})
were directly simulated by dint of the Runge-Kutta procedure of the sixth
order, \cite{mi1D}, to verify stability properties predicted by the linear
analysis. This was done by performing direct simulations with generic small
random perturbations added to the initial conditions.

\subsection{The dispersion relation}

The search for localized solutions -- here, complexes formed by discrete
solitons in each BEC component -- starts from the derivation of the
underlying dispersion relations. To this end, we substitute $\psi
_{n}^{+}=\psi ^{+}\exp {(i(kn-\mu t))}$ and $\psi _{n}^{-}=\psi ^{-}\exp {%
(i(kn-\mu t))}$, with wavenumber $k$, into the linearized version of Eqs. (%
\ref{eq2a}) and (\ref{eq2b}). Straightforward algebra gives rise to the
following dispersion relation for both intra-SOC and inter-SOC systems:
\begin{equation}
\mu _{1,2}=-2C\cos k\pm \sqrt{\Omega ^{2}+4\kappa ^{2}\sin ^{2}k}.
\label{dispersion}
\end{equation}%
Further analysis of the dispersion relation is worded in terms of the
intra-SOC model, for which discrete solitons are actually produced below.

\begin{figure}[h]
\centering {\includegraphics[width=9cm]{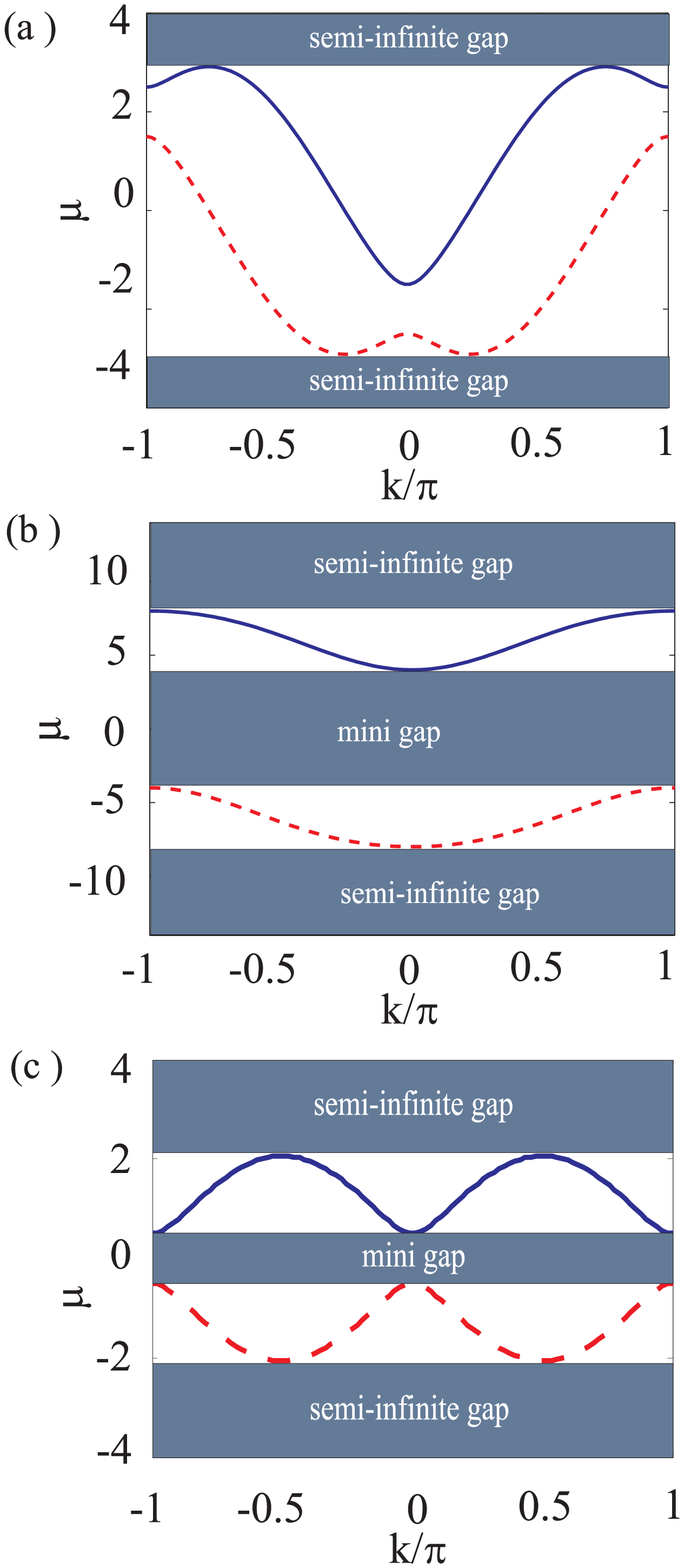}}
\caption{(Color online) Dispersion curves, as given by Eq. (\protect\ref%
{dispersion}), for (a) $\Omega =-0.5$, $C=1$; (b) $\Omega =-6$, $C=1$; and
(c) $\Omega =-0.5$, $C=0$. In all the plots, $\,\protect\kappa =1$. At $%
|\Omega |>2C$, the mini-gap Eq. (\protect\ref{minigap}) opens up around $%
\protect\mu =0$. This figure corresponds to both systems considered here, of
the intra-SOC and inter-SOC types, i.e., given by Eqs. (\protect\ref{eq2a})
and (\protect\ref{eq2b}), respectively.}
\label{fig1}
\end{figure}

In the absence of the linear coupling between the two components, i.e., $%
\Omega =0$, Eq. (\ref{dispersion}) reduces to a straightforward
generalization of the commonly known spectrum of the discrete linear Schr%
\"{o}dinger equation. In the present case, the spectrum corresponding to $%
\Omega =0$ contains the Bloch band, $\left\vert \mu _{1,2}\right\vert \leq 2%
\sqrt{C^{2}+\kappa ^{2}}$, sandwiched between two semi-infinite gaps. As
follows from elementary analysis of Eq. (\ref{dispersion}), and shown in
Fig. \ref{fig1}(a), the same remains qualitatively correct if the SOC is
relatively weak, \textit{viz}., at $|\Omega |\leq 2C$. Further increase of
the strength of the linear coupling between the components opens a \textit{%
mini-gap} in the middle of the Bloch band, which exists at $|\Omega |>2C$,
occupying the region of
\begin{equation}
\left\vert \mu _{1,2}\right\vert <\left\vert \Omega \right\vert -2C,
\label{minigap}
\end{equation}%
as shown in Fig. \ref{fig1}(b). Usually, mini-gaps open up in the spectrum
of 1D \cite{mini} and 2D \cite{we} \textit{superlattices}, i.e., discrete
lattices which are subjected to an additional spatially periodic modulation.
A mini-gap is also present in the spectrum of the discrete coupler
introduced in Ref. \cite{kgm}. In the present case, it may be understood,
roughly, as a result of the precession of the pseudo-spin around the
magnetic axis. Lastly, Fig. \ref{fig1}(c) displays the limit case of $C=0$,
when the usual inter-site hoppings are negligible in comparison with the
SOC-induced linear inter-site coupling, while the semi-infinite bandgaps and
the mini-gap are still present in the spectrum.

The tight-binding approximation underlying the derivation of the discrete
equations takes into account only the first Bloch band, while the
higher-order ones are neglected. This means that the band is sandwiched, as
said above, between two semi-infinite bandgaps [see Fig. \ref{fig1}(a)]. In
the presence of the attractive nonlinearity, the existence of fundamental (%
\textit{unstaggered}) discrete solitons is expected in the lower
semi-infinite gap. Solitons of this type are chiefly considered in this
work. On the other hand, staggered solitons, with $\pi $-shifted fields at
adjacent sites, may exist in the mini-gap \cite{we,kgm,PR463}. Basic results
for the latter type of the discrete solitons are briefly reported below too.

\section{Results and discussion}

In the following, results are chiefly presented for the intra-SOC model,
based on Eq. (\ref{eq2a}). Qualitatively similar conclusions can be obtained
for the inter-SOC model based on (\ref{eq2b}), which is explained by the
fact that lattice solitons tend to be similar in different systems, due to
the strong effect of the discretization. Recently, both 2D and 1D discrete
models with the inter-SOC terms (the Rashba coupling) were considered in
\cite{Fukuoka2}; however, unlike the present analysis, which focuses on
strongly discrete states, that work was dealing with broad discrete solitons
corresponding to the quasi-continuum regime.

\begin{figure}[h]
\centering {\includegraphics[width=15cm]{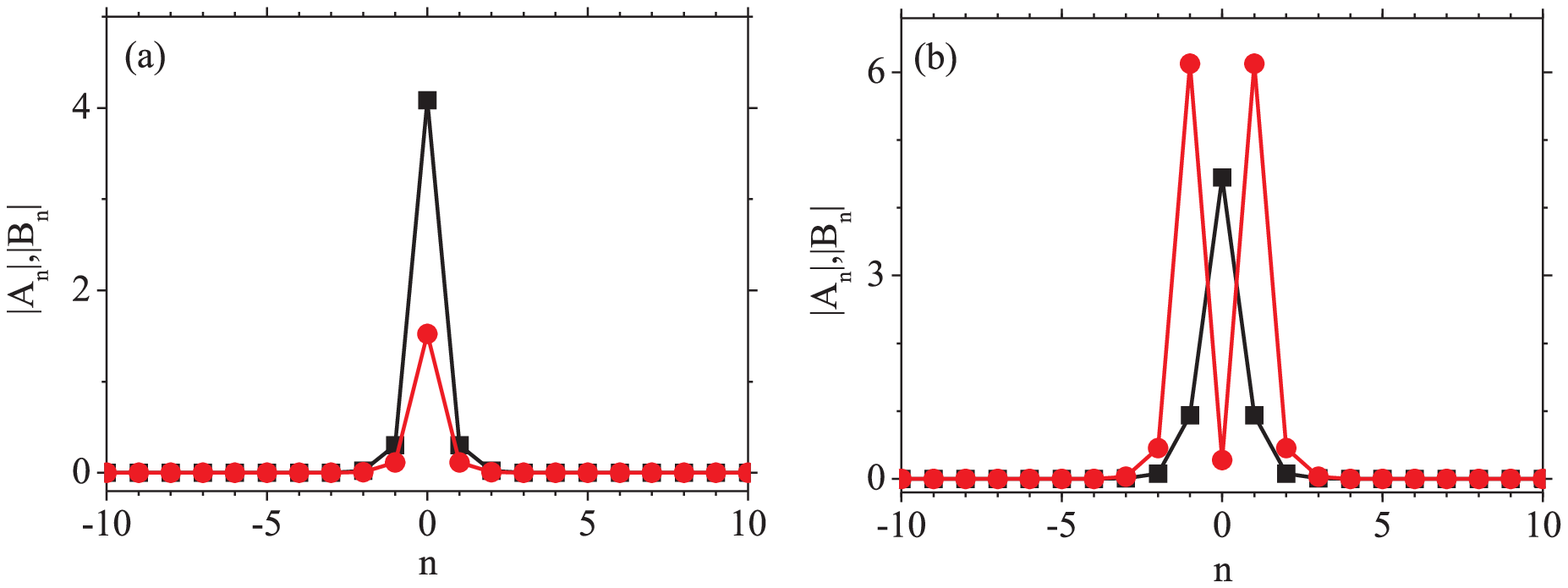}}
\caption{(Color online) Panels (a) and (b) displaye typical shapes of the
discrete miscible and immiscible complexes (MCs and IMCs), respectively.
Black (squares) and red (circles) lines denote, severally, components $A_{n}$
and $B_{n}$. The respective parameters in Eq. (\protect\ref{eq2as}) are $%
\protect\kappa =1$, $\,\protect\gamma _{1}=-1$,$\,\ \protect\gamma =-0.5$,$%
\,\ \protect\beta =-1$,$\,\ \Omega =-2$, $\protect\mu =-19.4,\,C=1$ .}
\label{fig2}
\end{figure}

The discrete SOC-BEC systems support different types of bright solitons,
depending on the systems' parameters. They are classified here as miscible
and immiscible complexes (MC and IMC, respectively, see Fig. \ref{fig2}),
formed of localized patterns in both components \cite{transmalomed,miIMT}.
We consider as miscible the localized complexes in which both stationary
components ($A_{n}$ and $B_{n}$) are overlapping, see Fig. \ref{fig2}(a). On
the other hand, immiscible states are characterized by separated shapes of
the two components. The numerical results demonstrate that stable IMCs are
mostly formed by one component featuring two maxima and the other component
with a single maximum between these two, see Fig. \ref{fig2}(b). It is
relevant to mention that broad (quasi-continual) discrete solitons found for
the inter-SOC system in \cite{Fukuoka2} also feature miscible and immiscible
structures, which are called, in that context, ``smooth" and ``striped" ones
(the latter pattern demonstrates a large number of alternating local peaks
of the density of the two components).

Our analysis has identified four basic types of the MC solutions, which may
be built as follows (cf. Ref. \cite{miPRB}): in-phase on-site solitons
(IPOS), $\mathrm{Re}\left( A_{n}\right) \times \mathrm{Re}\left(
B_{n}\right) >0$, counter-phase on-site solitons (CPOS), $\mathrm{Re}\left(
A_{n}\right) \times \mathrm{Re}\left( B_{n}\right) <0$, in-phase inter-site
solitons (IPIS), $\mathrm{Re}\left( A_{n}\right) \times \mathrm{Re}\left(
B_{n}\right) >0$, and counter-phase inter-site solitons, $\mathrm{Re}\left(
A_{n}\right) \times \mathrm{Re}\left( B_{n}\right) <0$ (CPIS), see Fig. \ref%
{fig3}.

\begin{figure}[h]
\centering {\includegraphics[width=12cm]{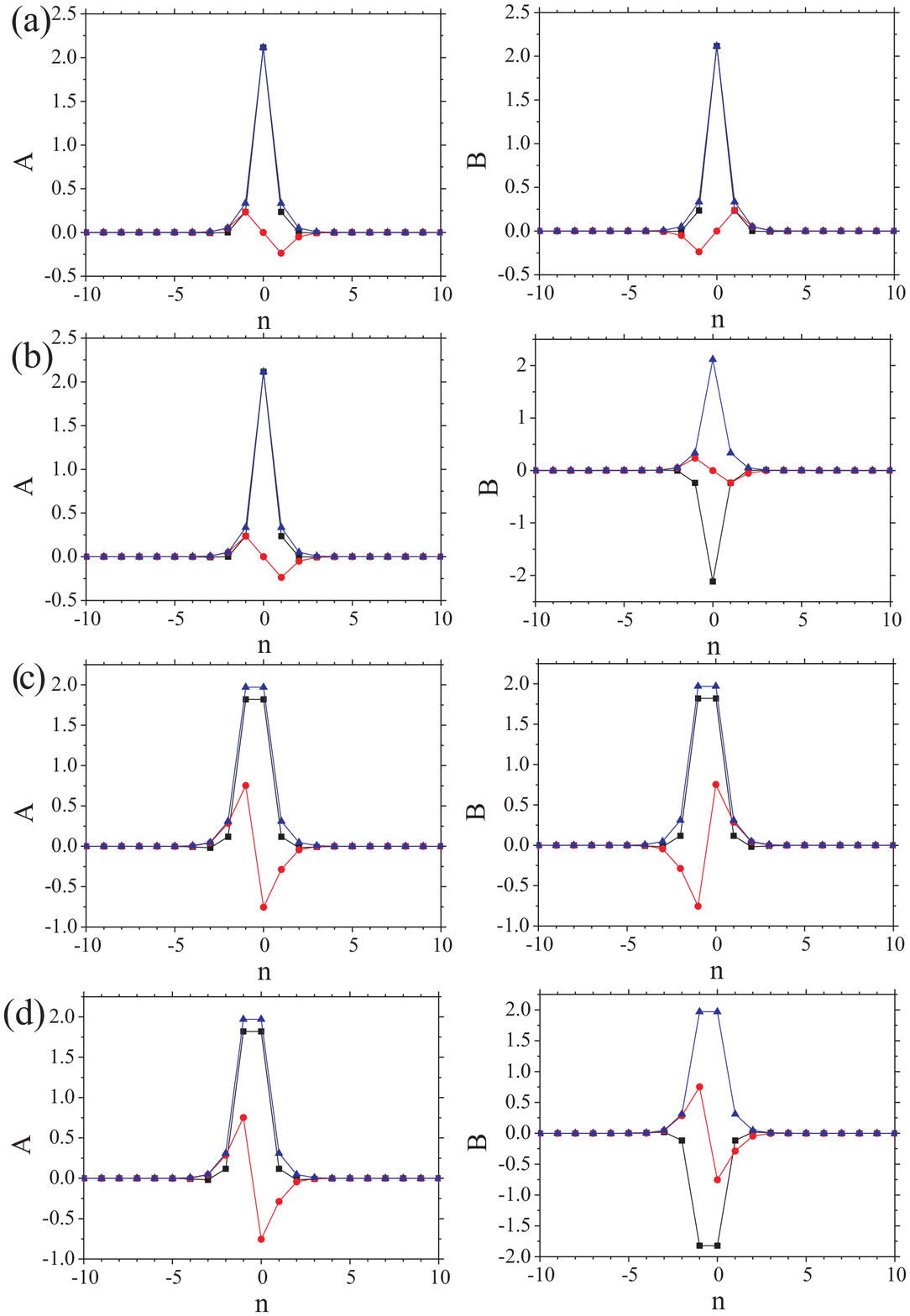}}
\caption{(Color online) Examples of discrete two-component solitons of the
miscible type: (a) an IPOS complex, (b) a CPOS complex, (c) an IPIS complex,
and (d) a CPIS complex. Black (squares), red (circles), and blue (triangles)
lines depict real and imaginary parts and the absolute value of the fields,
respectively. The parameters are $\protect\kappa =1,\,\protect\gamma =%
\protect\gamma _{1}=-1,\,C=1,\,\protect\mu =-9.4$.}
\label{fig3}
\end{figure}

Considering the IMCs, various structures of the on- and inter-site types
have been found, which reflects strong sensitivity of the system to
parameter values. We report here only those IMCs which may be stable, and
mention scenarios of the transformation of unstable IMCs into MCs. In fact,
only on-site IMCs are found to be stable in certain ranges of the parameter
space.

Miscible on-site complexes of both the in-phase and counter-phase types are
found to be stable over large parts of their existence region in parameter
space ($\gamma ,\beta ,\kappa ,\Omega ,C$). As mentioned above, they belong
to the lower semi-infinite gap of the systems' spectra, shown in Fig. \ref%
{fig1}. IPOS complexes lose the stability only in a limited parameter area;
for example, in the case of $C=1$ they are unstable at $\beta >-1$ and $%
-2<\Omega <-0.5$. In direct simulations, the instability in that area leads
to spontaneous formation of breathing localized structures with a smaller
norm, in comparison with the original unstable discrete solitons. Without
the inter-species attraction, $\beta =0$, IPOS solutions are unstable,
except when the Rabi coupling is absent too, $\Omega =0$, making the system
decoupled. On the other hand, CPOS are unstable at $\beta <-1$ and $\Omega
\neq 0$. Without the attraction between the components, $\beta =0$, and for
small values of the Rabi coupling parameter ($-1<\Omega <0$), CPOS are
stable. For higher values of $|\Omega |$, CPOS solutions break into
breathing structures. Both components of the complex exhibit the same
behavior and evolve into miscible breathers, exchanging the power (norm) in
the course of the evolution. All other miscible solutions were found to be
unstable.

\begin{figure}[h]
\centering {\includegraphics[width=12cm]{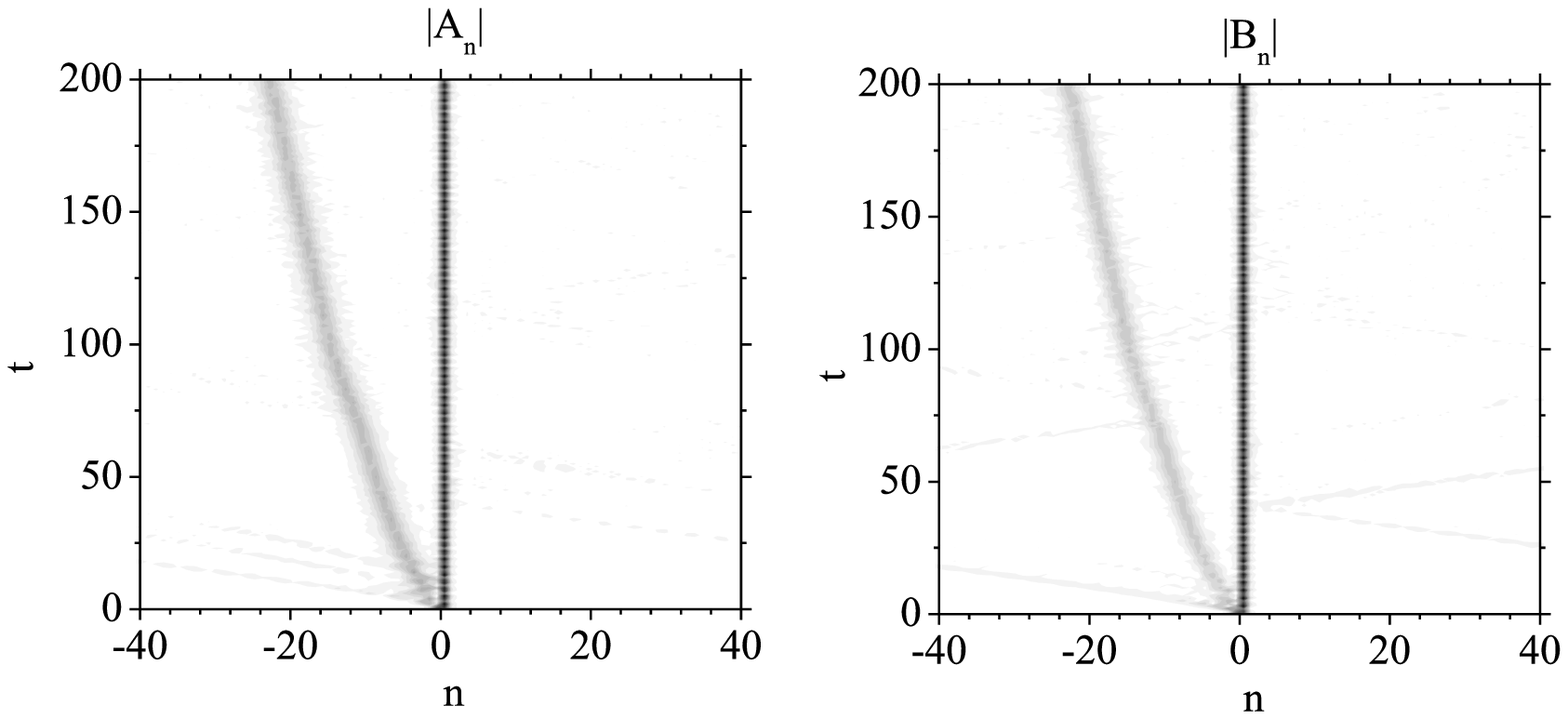}}
\caption{The evolution of two components of an inter-site CPIS mode with $%
\protect\mu =-9.4$, $\protect\kappa =1$, $\protect\gamma =-0.5$, $\protect%
\gamma _{1}=-1$, $\protect\beta =-2$,$\,\Omega =-0.5$, and $C=1$. Both
components break into a trapped on-site mode and a moving one. Initial small
random perturbations are added to the solutions to catalyze the onset of the
instability.}
\label{fig4}
\end{figure}

All inter-site-centered solitons are found to be unstable, as it usually
happens in other discrete systems. A trend of the evolution of these
unstable modes revealed by the simulations is to end up as on-site breathers
with significant emission of matter waves. In particular, the CPIS complexes
evolve so that both components exhibit the same behavior and evolve into
miscible on-site breathing structures exchanging the power in the course of
the evolution. In this case, a part of the emitted matter wave propagates in
the form of a new two-component moving complex, see Fig. \ref{fig4}.

Generally, we conclude that all unstable soliton solutions tune their
energy, by radiating away a part of it, to the energy corresponding to the
miscible on-site solitons existing at the same parameters . The modes evolve
as stable on-site breathers. If the radiated energy, transferred to the
background, is large enough, as is the case for the CPIS mentioned in the
previous paragraph, new moving localized breathing structures can be formed
(Fig. \ref{fig4}).

The IMC solutions tend to get destabilized with the increase of the strength
of the linear coupling between the components. Usually, they too evolve into
miscible breathing modes after radiating away a part of the energy. Stable
IMCs feature on-site two-component configurations, see Fig. \ref{fig2}(b).%
\textbf{\ }In the area of the system's parameters where the IMCs are stable
in the inter-SOC systems, a stable branch of MCs of the on-site type can be
found too, which implies bistability between the IMC and MC modes.\textbf{\ }%
Then, selection of one of the coexisting solutions is determined by initial
conditions. By checking values of the Hamiltonian, we have found that, in
the bistability region (the whole gray area in Fig. \ref{fig6} (b)), the MCs
feature smaller energy, hence they realize the ground state of the system.
The energy difference between IMC and MC in this case is smaller for smaller
values of $C$. Similar bistability is found in the inter-SOC system for the
parameter set shown in Fig. \ref{fig6}(a) in a narrow area with very small $%
\Omega \approx 0$ and $0<C<0.2$.

\begin{figure}[h]
\centering {\includegraphics[width=13cm]{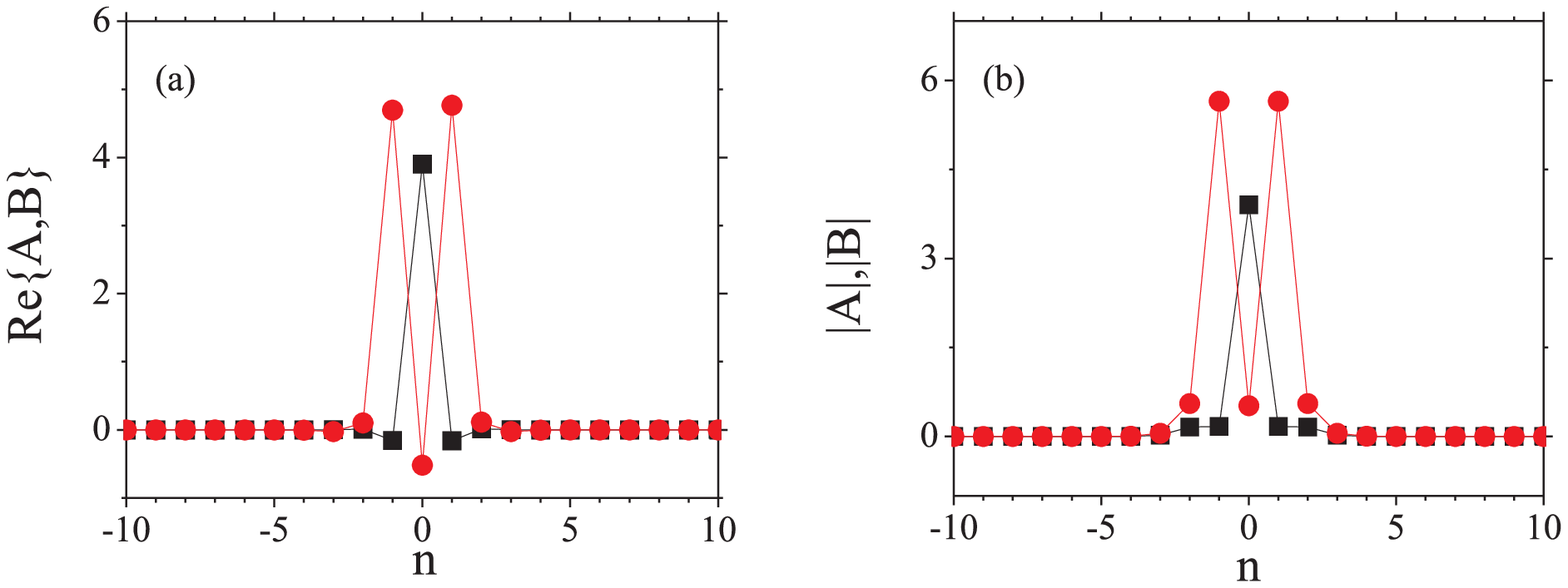}}
\caption{(Color online) An example of a stable IMC (immiscible complex): (a)
real parts of its two components; (b) the corresponding absolute values. The
parameters are: $\Omega =-4$,\thinspace $\ \protect\gamma _{1}=-1$%
,\thinspace $\ \protect\gamma =-0.5$,$\,\ \protect\beta =-5$,$\,\ \protect%
\mu =-16$,$\,\ \protect\kappa =1$ and $C=1$. Black lines with squares and
red lines with circles are associated with components $A$ and $B$,
respectively. }
\label{fig5}
\end{figure}

Unlike the intra-SOC system, in its inter-SOC counterpart the SOC can induce
destabilization of the IMCs (leading to a spontaneous transition from the
IMC to an MC) if one sets $\beta =\Omega =C=0$ in Eq. (\ref{eq2b}), so that
the two components interact solely via the linear SOC terms $\sim \kappa $.
This happens in a narrow area of the remaining parameter space.

On the other hand, the SOC affects the threshold of the
immiscibility-miscibility transition in both the intra-SOC and inter-SOC
systems, in cases of the nonlinear coupling only ($\Omega =0$), linear
coupling only ($\beta =0$), and when both couplings are present ($\Omega
\neq 0,\beta \neq 0$). Figure \ref{fig6} displays the position of the
miscibility threshold, i.e., the curve separating the region of the
coexistence of stable IMC and MC, and the region where the IMC is unstable
while the MC is stable (as concerns the on-site configurations), in the $%
(|\Omega |,C)$ parameter plane. In the inter-SOC system [Fig. \ref{fig6}%
(a)], parameter $\beta =-1$ and $\kappa =1$ are fixed. In the intra-SOC
system, we set $\beta =0$ and $\kappa =1$, because for $\beta <-0.5$ stable
IMC have not been found [Fig. \ref{fig6}(b)].

\begin{figure}[h]
\centering {\includegraphics[width=12cm]{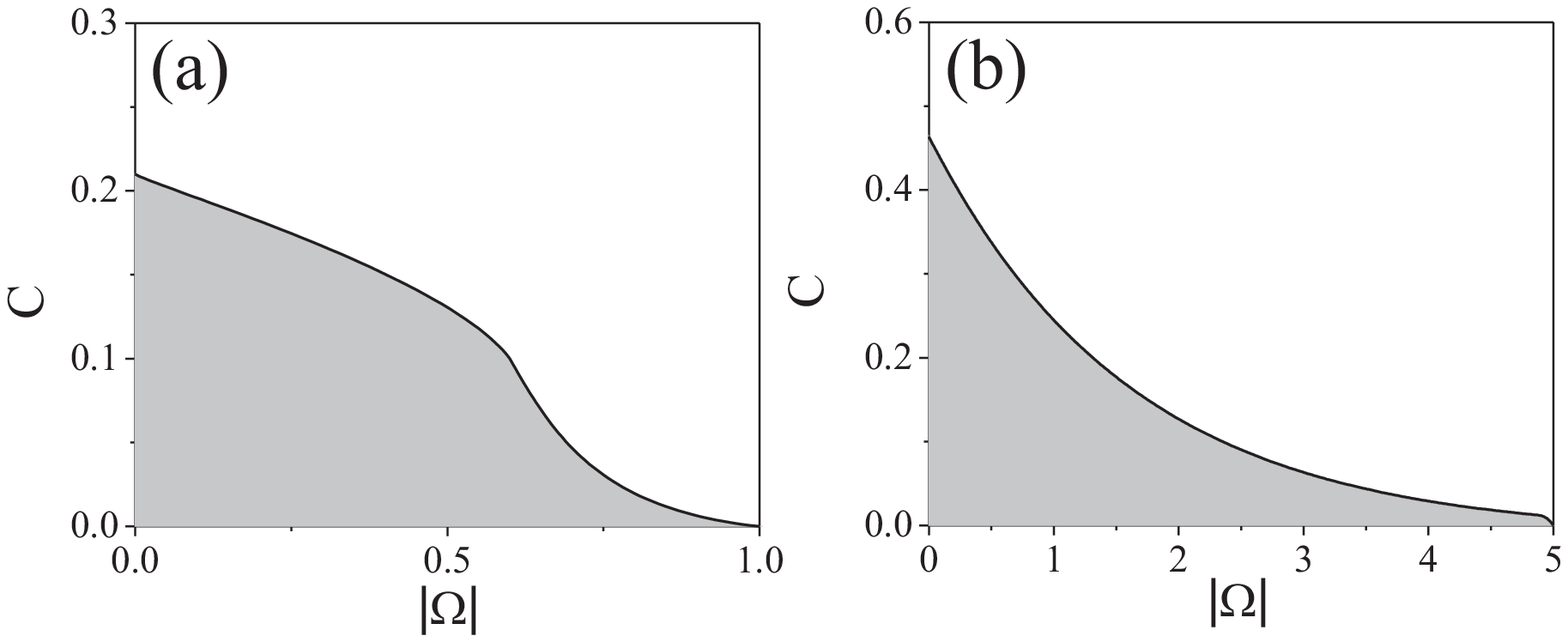}}
\caption{(Color online) The transition from stable (gray areas) to unstable
(white areas) IMC in the $(|\Omega |,C)$ plane: (a) in the inter-SOC system
[Eq. (\protect\ref{eq2b}) with $\protect\beta =-1$], and (b) in the
intra-SOC one [Eq. (\protect\ref{eq2a}), with $\protect\beta =0$]. Other
parameters are $\protect\gamma =\protect\gamma _{1}=-1$, and $\protect\kappa %
=1$. In the area with small $|\Omega|$ in the grey area in plot
(a) stable IMC and MC can exist. While in the plot (b) in the
entire gray area both stable IMCs and MCs can be created
(bistability regions). Those IMCs which are unstable (white areas
in figure) spontaneously evolve into stable MC breathers, after
radiating away a part of the initial norm.} \label{fig6}
\end{figure}

Figure \ref{fig6} demonstrates that the IMC complexes have the largest
stability region in the case of $C=0$, when the usual hopping is negligible,
while the SOC terms are present [recall that, as shown in Fig. \ref{fig1}%
(c), the respective linear spectrum remains meaningful, always featuring the
mini-gap]. This case corresponds to the setting when the OL potential is
very deep, and, simultaneously, the SOC, induced by the laser beams
illuminating the condensate, is strong. Such a setting is feasible, because,
in the underlying tight-binding approximation, the hopping coefficient, $C$
in Eqs. (\ref{eq2a}) and (\ref{eq2b}), is proportional to the square of the
small overlap integral of wave functions (Wannier modes \cite{BECinOL})
localized at adjacent sites of the OL, while the scaled SOC coefficient, $%
\kappa $, is proportional to the first power of the same integral. In fact,
this limit case emulates the spin-orbit-coupled binary condensate made of%
\emph{\ infinitely heavy atoms} (as mentioned above, the case of an infinite
particle's mass was recently considered in \cite{Basque}, in the context of
a continual model elaborated for the measurement of spin).

Furthermore, in the same limit of $C\rightarrow 0$, the staggering
transformation in the form of $\psi _{n}=i^{n}\tilde{\psi}_{n}$ \cite{kgm},
cf. Eq. (\ref{stag}), can be applied to the intra-SOC system, transforming
it into a discrete system without the SOC terms, but, effectively, with the
hopping coefficients of opposite signs, $\kappa $ and $-\kappa $, in the two
components:%
\begin{eqnarray}
i\frac{\partial \tilde{\psi}_{n}^{+}}{\partial t} &=&\kappa (\tilde{\psi}%
_{n+1}^{+}+\tilde{\psi}_{n-1}^{+})+(\gamma _{1}|\tilde{\psi}%
_{n}^{+}|^{2}+\beta |\tilde{\psi}_{n}^{-}|^{2})\tilde{\psi}_{n}^{+}-\Omega
\tilde{\psi}_{n}^{-}~,  \nonumber \\
i\frac{\partial \tilde{\psi}_{n}^{-}}{\partial t} &=&-\kappa (\tilde{\psi}%
_{n+1}^{-}+\tilde{\psi}_{n-1}^{-})+(\beta |\tilde{\psi}_{n}^{+}|^{2}+\gamma |%
\tilde{\psi}_{n}^{-}|^{2})\tilde{\psi}_{n}^{-}-\Omega \tilde{\psi}_{n}^{+}~.
\label{eq2aa}
\end{eqnarray}%
In this way, the model can emulate a two-component BEC with \emph{opposite
signs of the effective atomic masses}, which is not possible in the
implementation of the BEC.

\begin{figure}[h]
\centering {\includegraphics[width=12cm]{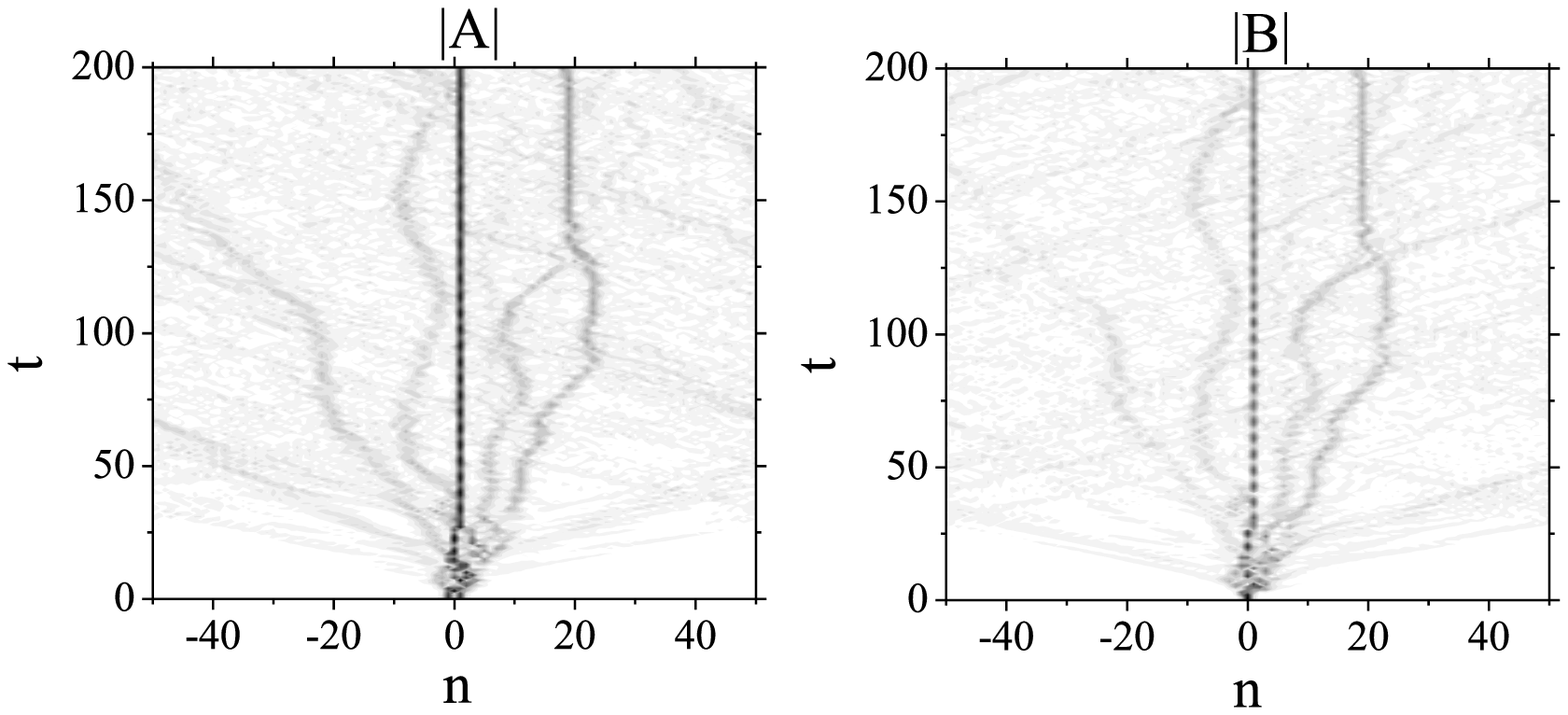}}
\caption{The evolution of an unstable immiscible complex (IMC) in the
intra-SOC system with $\protect\mu =-19.4$, $\protect\kappa =1$, $\,\protect%
\gamma =-0.5$, $\protect\gamma _{1}=-1$,$\,\ \protect\beta =-1$,$\,\Omega
=-2 $, and $C=1$. Initial small random perturbations were added to the
immiscible localized complex.}
\label{fig7}
\end{figure}

Continuing the consideration of IMC modes in the intra-SOC system, we
outline typical instability-development scenarios for them (similar
scenarios are produced by the inter-SOC system). Unstable on-site IMCs may
actually be very robust overall, evolving into localized breathing miscible
structures, see Fig. \ref{fig7}. In that case, the two components of the
on-site IMC evolve into miscible breathing structures, which exchange the
norm in the course of the evolution.

\begin{figure}[h]
\centering {\includegraphics[width=12cm]{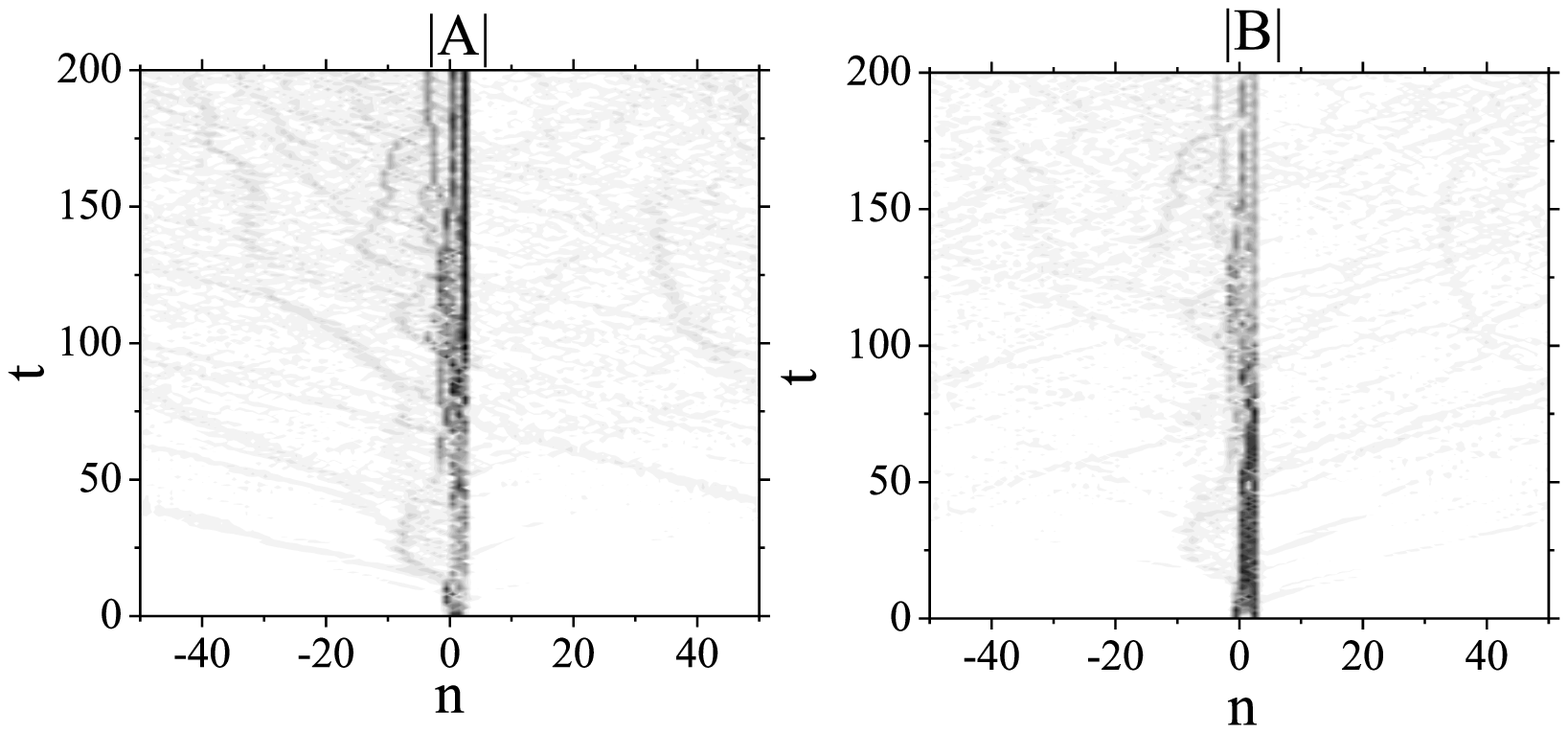}}
\caption{(Color online) The evolution of an unstable inter-site IMC in the
intra-SOC system. The parameters are the same as in Fig. \protect\ref{fig7}.
}
\label{fig8}
\end{figure}

Figure \ref{fig8} illustrates the dynamics of an unstable inter-site IMC. In
the course of the evolution, a large share of the norm is radiated away, and
the formation of irregular localized patterns is observed in a neighborhood
of the initial position of the components.

All the localized complexes considered above are generated in the
semi-infinite gap, in terms of the spectra of the linearized systems. It
remains to briefly describe what happens when a mini-gap opens at $|\Omega
|>2C$, as shown in Figs. \ref{fig1}(b) and (c). In that gap, we have found
miscible CPOS and CPIS complexes, which resemble similar complexes found in
the semi-infinite gap, see Fig. \ref{fig3}. Only the CPOS complexes may be
stable in certain parameter areas inside the mini-gap. All other structures
are highly unstable in this case.

To summarize this section, the stable on-site localized MCs (miscible
complexes) can be created in the large area of the parameter space in the
two-component SOC-BEC system. On the other hand, stable IMCs (immiscible
complexes) were found only in a narrow parametric area. The latter one is
often a bistability region, where the system supports both the IMC and MC
stable complexes (which is usually not possible in the free space \cite%
{Fukuoka2}). Unstable IMCs, as well as unstable inter-site MCs, evolve into
breathing miscible states, emitting matter waves away. These conclusions
pertain to both the intra-SOC and inter-SOC systems considered. Thus, we
conclude that, in general, all inter-species couplings, linear (SOC) and
attractive nonlinear ones, favor miscible states.

\section{Conclusions}

We have introduced a discrete one-dimensional model of SOC\
(spin-orbit-coupled) binary BEC trapped in a deep OL potential. Two systems
were considered, of the intra-SOC and inter-SOC types, in which the SOC
terms, represented by the discrete version of the first spatial derivatives,
act, respectively, inside of each component corresponding to the dressed
atomic state, or couple the different components. Each system includes
attractive intra- and inter-species cubic interactions (repulsive
interactions can be transformed into the same form by means of staggering).
It has been demonstrated that both systems support two distinct types of
discrete soliton complexes, miscible and immiscible ones, depending on
values of the systems' parameters. The transition between the two types of
the complexes can be controlled by the SOC\ strength. As usual, only
on-site-centered modes may be stable. The SOC terms may open up a mini-gap
in the systems' spectrum, where stable miscible on-site soliton complexes
exist too.

New possibilities offered by these settings are scenarios for the
emulation of SOC binary condensates built of infinitely heavy
atoms, as well as of the binary BEC with effective atomic masses
which have opposite signs. Both settings, that correspond to the
limit when the usual hoppings are negligible, while the SOC keeps
acting, are obviously unavailable in the direct realization of
BEC.

To continue the present analysis, it may be interesting to consider bound
states of the discrete solitons, as well as their mobility. A challenging
problem is to extend the analysis to the 2D setting, where it has been
recently demonstrated that the SOC, in the combination with the nonlinear
attraction,\ supports stable bright solitons in the in the presence of the
OL potential \cite{Konotop,Fukuoka2}, as well as in\emph{\ free space} \cite%
{Fukuoka,Fukuoka2} (without the SOC, all 2D solitons sustained by the cubic
attractive interaction in the free space are unstable against the critical
collapse).

\section*{Acknowledgments}

P.P.B., G.G., J.P., A.M., and Lj.H. acknowledge support from the Ministry of
Education and Science of Serbia (Project III45010). B.A.M. appreciates
valuable discussions with H. Sakaguchi, V. V. Konotop, Y. V. Kartashov, and
L. Salasnich.

\end{document}